\documentclass[aps, prl, reprint, superscriptaddress]{revtex4-2}
\usepackage{amsmath, newtxtext, newtxmath}
\usepackage{graphicx}
\usepackage{dcolumn}
\usepackage{bm}
\usepackage{soul}
\usepackage{hyperref}
\hypersetup{
    colorlinks=true,        
    linkcolor=blue,         
    citecolor=blue,        
    filecolor=magenta,      
    urlcolor=blue           
}
\usepackage{color}
\newcommand{\textbl}{\textcolor{black}}
\newcommand{\textmg}{\textcolor{black}}
\usepackage{braket}

\begin{document}

\title{Two Spin-State Crystallizations in LaCoO$_{3}$}


\author{Akihiko~Ikeda}
\email[]{ikeda@issp.u-tokyo.ac.jp}
\author{Yasuhiro~H.~Matsuda}
\email[]{ymatsuda@issp.u-tokyo.ac.jp}
\affiliation{Institute for Solid State Physics, University of Tokyo, Kashiwa, Chiba 277-8581, Japan}
\author{Keisuke~Sato}
\affiliation{National Institute of Technology, Ibaraki College, Hitachinaka, Ibaraki 312-0011, Japan}

\date{\today}

\begin{abstract}

We report a magnetostriction study of a perovskite $\rm{LaCoO}_{3}$ above 100 T using our state-of-the-art strain gauge to investigate an interplay between electron correlation and spin crossover.
There has been a controversy regarding whether two novel phases in $\rm{LaCoO}_{3}$ at high magnetic fields result from crystallizations or Bose-Einstein condensation during spin crossover as manifestations of localization and delocalization in spin states, respectively.
We show that both phases are crystallizations rather than condensations, and  the two crystallizations are different, based on the observations that the two phases exhibit as magnetostriction plateaux with distinct heights.
\textbl{The crystallizations of spin states have emerged manifesting the localizations and interactions in spin crossover with large and cooperative lattice changes.}

\end{abstract}

\maketitle

\textmg{
In condensed matters, novel phases often manifest themselves at boundary regions between different competing phases.
Transition metal oxide is one of the most abundant field, where such emergent quantum states as unconventional superconductivity, charge-orbital order, spin density wave and magnetic orders emerge,  leading further to the functional phenomena such as multifferoic transitions, colossal magneto-resistance, and metal-insulator transition, representatively observed in cuprates and manganites \cite{cuprate, Tokura}.
The rich variety of the abovementioned phenomena is a consequence of the competing nature of the wave-particle duality of electrons due to electron correlation, with the coupled multiple degree of freedoms such as charge, orbital, spin, and lattices \cite{ImadaRMP}.
}

\textmg{
Among transition metal oxides, cobaltites are characterized by the occurrence of a spin crossover between high-spin (HS) and low-spin (LS) states due to the balance of Hund's coupling and crystal field splitting, that serves as a new degree of freedom in the correlated electron systems \cite{Hund}.
Therefore, our prime interest is to investigate the emergence of novel phases in cobaltites at proximity to a spin crossover.
It deepens the understanding of cobaltites, and is more generally essential to the correlated electron physics with multiorbitals such as, iron-based unconventional superconductivity, and Kondo physics in heavy fermion systems and excitonic condensation \cite{KunesJPCM}.
Besides the electronic states are associated to vast phenomena in cobalt-based materials such as metal-organic magnets \cite{Organic}, spin crossover, superconductivity \cite{Takada}, large thermoelectric phenomena \cite{Terasaki}, Lithium ion batteries \cite{LIB}.
}

 \begin{figure}[b]
 \includegraphics[width =\linewidth]{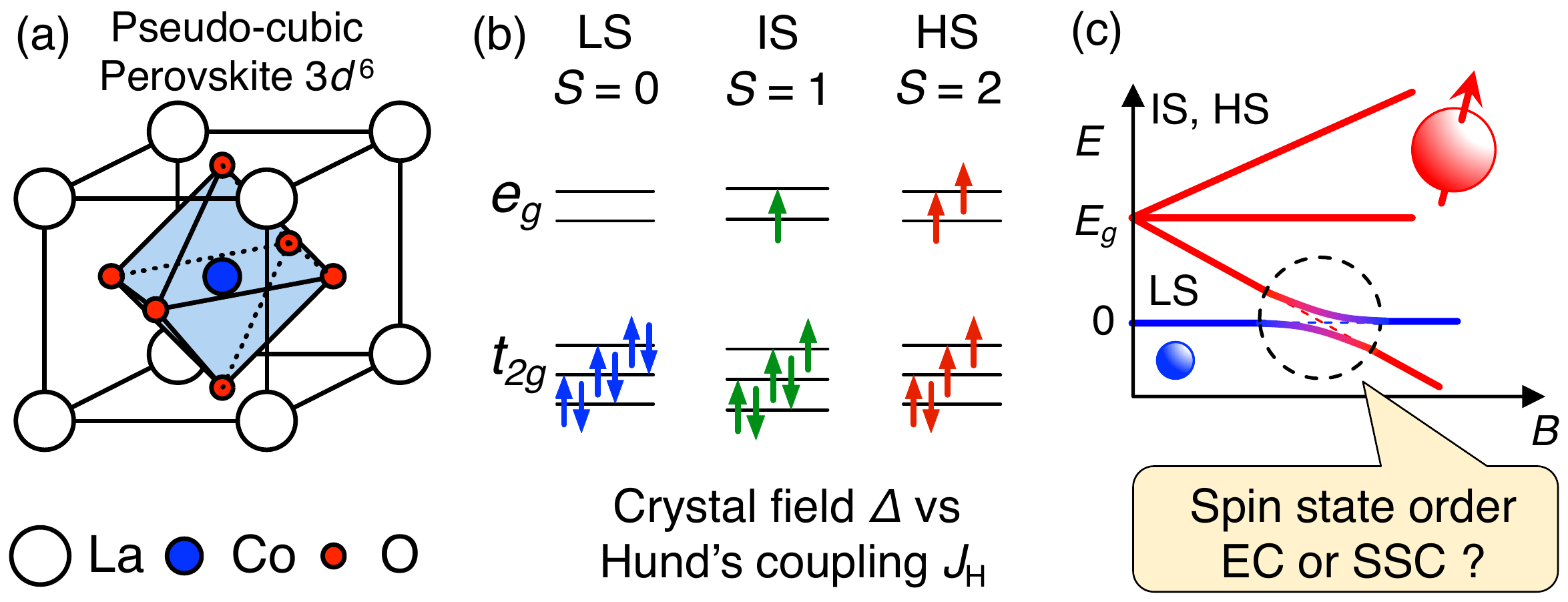}%
 \caption{(a) The lattice model of a pseudo cubic $\rm{LaCoO}_{3}$.
 (b) Electron and spin configurations in LS, IS and HS states.
 (c) A schematic diagram of a magnetic field induced spin crossover. 
 \textbl{EC and SSC stands for excitonic condensation and spin-state crystallization, respectively.} \label{idea}}
 \end{figure}

\textmg{
A perovskite cobaltite LaCoO$_{3}$ is the most prominent compound suitable for hunting emergent phases at proximity to a spin crossover because the spin crossover is controllable by hole doping \cite{Goodenough}, temperature \cite{Raccah}, tensile strain \cite{FujiokaPRL2013, FujiokaPRB2015, YokoyamaPRL2018} and magnetic fields \cite{SatoJPSJ2009, MoazPRL2012}.
The temperature evolution has stimulated many controversies among researchers. The change from the low-lying non-magnetic LS insulator ($t_{2g}^{6}$, $S=0$) state to a paramagnetic insulator state at 100 K and then to paramagnetic metal phase above 500 K \cite{Goodenough, Raccah} has been causing an intriguing debate over the possible occupation of the HS ($t_{2g}^{4}e_{g}^{2}$, $S=2$) state \cite{Noguchi, Ropka, Phelan, Podlesnyaks, Haverkort, Knizek, Mukhopadhyay} or the intermediate spin (IS, $t_{2g}^{5}e_{g}^{1}$, $S=1$) state \cite{Korotin, Mizokawa, Saitoh, Yamaguchi, Ishikawa} [Figs. \ref{idea}(a) and \ref{idea}(b)].
}

%
 \begin{figure*}
 \includegraphics[width = \linewidth]{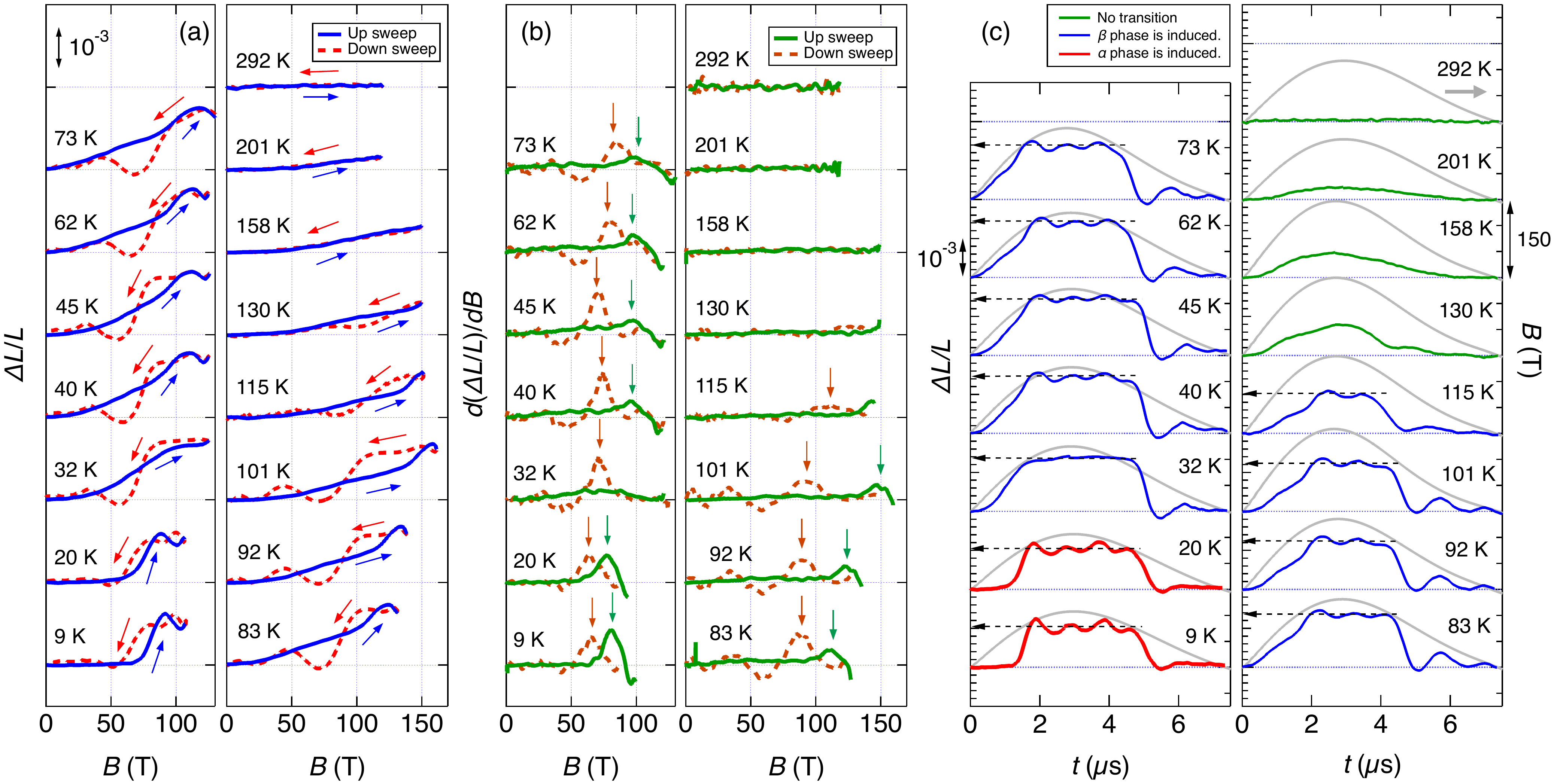}%
 \caption{\textbl{Magnetostriction curves of polycrystalline $\rm{LaCoO}_{3}$ up to 160~T measured by means of the fiber Bragg grating based high-speed 100 MHz strain gauge \cite{AIkedaFBG2017} at various temperatures from 9 to 292~K.
  (a) $\Delta L /L$ as a function of magnetic fields.
  Blue curves and dotted red curves are data from up and down sweep of magnetic fields, respectively.
  (b) $d(\Delta L /L)/dB$ as a function of magnetic fields.
  Green and dotted brown curves are data from up and down sweep of magnetic fields, respectively.
  The green and brown vertical arrows point to the peaks of $d(\Delta L /L)/dB$ data, which represent the critical fields for the spin-state transitions to $\alpha$ and $\beta$ phases.
  (c) $\Delta L /L$ curves and the external magnetic fields  (grey curves) shown as a function of time.
  Red and blue curves of $\Delta L /L$ correspond to the temperature ranges where the spin-state transition to $\alpha$ and $\beta$ phases take place, respectively.
  Green curves of $\Delta L /L$ correspond to the temperature range where no transition is observed. }
 \label{results}}
 \end{figure*}

\textmg{
Among them, the magnetic field effect on LaCoO$_{3}$ is the most controversial over the origin of the two emergent phases observed at very high magnetic fields above 100 T, being consistent with the large spin gap energy of $\sim 100 $ K [Fig. \ref{idea}(c)].
Below 30~K, a first order transition at 60~T is observed  \cite{SatoJPSJ2009, MoazPRL2012, RotterSR2014} with only a small increment of magnetization, $\approx 1/4$ of the saturation magnetization.
Above 30~K, the transition at 60~T disappears and another transition appears beyond 100~T \cite{AIkedaLCO2016}.
The high-field phases below and above 30 K are termed as $\alpha$ and $\beta$ phases, respectively.
As origins of $\alpha$ and $\beta$ phases, excitonic condensations described as $\ket{\rm{LS}+\rm{IS}}$, and HS-LS spin-state crystallizations are proposed theoretically \cite{TatsunoJPSJ, SotnikovSR}.
The wave-particle duality of HS or IS states is crucial to induce such non-trivial phases at the spin-crossover \cite{KunesJPCM}.
However, further experimental verification remains elusive.
One needs to observe the spin-state fractions as a function of magnetic fields for the experimental verification.
In analogy with the case of the dimer-spin systems, where Bose-Einstein condensation and a crystal of magnons exhibit a magnetization slope and a plateau, respectively \cite{Zapf}, in the correlated spin-crossover system, an excitonic condensation and a spin-state crystallization will exhibit a slope and a plateau of spin-state fractions, respectively, as is also theoretically suggested in the literatures \cite{TatsunoJPSJ, SotnikovSR}.
}

\textmg{
Magnetovolume effect is one of the most direct probe of spin-state fractions, owing to the firm couplings between  spin-states and the ionic volume.
HS and IS have larger volume than LS because of the increasing occupations of the extended $e_{g}$ orbitals [See Fig. \ref{idea}(c)].
We recently devised a state of the art high-speed strain gauge \cite{AIkedaFBG2017} to use it  well beyond 100 T, generated for a few $\mu$s duration with destructive pulse magnets.
Now, we can verify spin-state crystallizations and excitonic condensations in LaCoO$_{3}$ well above 100 T via the magneto-lattice couplings.
}

\textbl{
In this paper, we report magnetostriction measurements of $\rm{LaCoO}_{3}$ up to well above 100~T using the new high-speed magnetostriction gauge \cite{AIkedaFBG2017}.
In both $\alpha$ and $\beta$ phases, $\Delta L/L$ show plateaux, where $\Delta L/L$ is $\approx 1.4$ times larger in $\beta$ phase than that in $\alpha$ phase.
The observations indicate that a spin-state crystallization is preferred over an excitonic condensation in each of the $\alpha$ and $\beta$ phases.
Origins of $\alpha$ and $\beta$ phases are discussed, considering IS and HS duality, magnetism and a recent calculation of the spin-state crystals \cite{TomiyasuArxiv}.
}

High magnetic fields up to 200~T are generated using a single-turn coil megagauss generator in Institute for Solid State Physics, Univ. of Tokyo, Japan \cite{Miura2003}, which is a destructive pulse magnet with a duration of 7 $\mu$s.
Magnetostriction measurements are performed by means of a high-speed strain gauge of 100 MHz, where an optical fiber with a fiber Bragg grating is directly glued onto rod-shaped polycrystalline and single crystalline samples of $\approx 2$ mm in length and $\approx1$ mm in diameter \cite{AIkedaFBG2017}.
All the magnetostriction data in the present paper are measured in the longitudinal direction, $\Delta L \parallel B$ \cite{note06}.

\begin{figure}[t]
 \includegraphics[width = \linewidth]{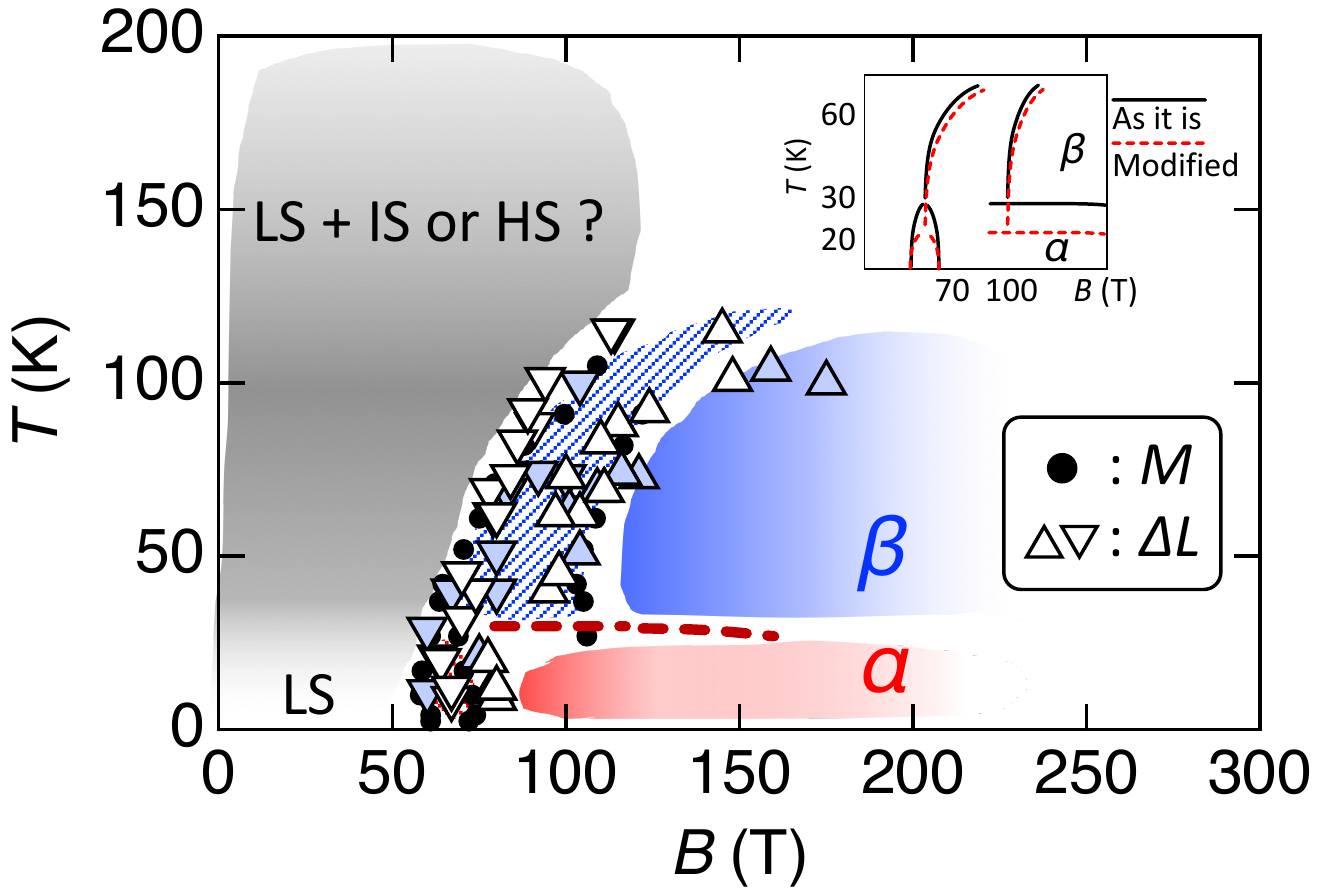}%
 \caption{
 \textbl{
 The obtained spin-state transition fields of $\rm{LaCoO}_{3}$ plotted on a $B$-$T$ plane.
  The triangles pointing up and down indicate the data in up sweep and down sweep, respectively.
  The triangles filled with white and pale blue color are data from polycrystalline [Fig. \ref{results}(b)] and single crystalline samples \cite{supp1}, respectively.
 Black filled circles are the previously obtained transition fields reported in Ref. \cite{AIkedaLCO2016} shown for comparison.}
\textmg{The inset shows a possible modification to the temperature reading of the obtained phase diagram, which is in demand considering the adiabatic condition of the sample during the $\mu$s pulsed magnetic field.
See main text for the details.
 }
 \label{pd}}
 \end{figure}

\textbl{Figs. \ref{results}(a)-\ref{results}(c) show the measured magnetostriction curves of polycrystalline $\rm{LaCoO}_{3}$ at various temperatures, where the sample and the strain gauge remained intact after each pulse.
\textmg{
$\Delta L/L$ data in Fig. \ref{results}(a) exhibit sudden increases of $\Delta L/L$ at high fields with large hystereses between the up-sweep (blue curves) and down-sweep (red dashed curves) data, indicating the first order spin-state transitions, being in good agreement with previous magnetization \cite{SatoJPSJ2009, MoazPRL2012}, magnetostriction studies \cite{MoazPRL2012, RotterSR2014} below 100 T, and a magnetization study beyond 100 T \cite{AIkedaLCO2016}.
}
The transition fields are identified with the vertical arrows in Fig. \ref{results}(b) pointing to the peaks of $d(\Delta L/L)/dB$ data, which are summarized on the $B$-$T$ plane in Fig. \ref{pd}, which shows a good agreement with the previous magnetization study beyond 100 T \cite{AIkedaLCO2016}.
Two striking features in Fig. \ref{pd} are that the phase boundaries shift to higher magnetic fields with increasing initial temperature and that the phase boundary with the up-sweep data shows discrete jump at the temperature of $\approx30$ K.
These indicate that $\beta$ and $\alpha$ phases are distinct ordered phases, as is speculated in Ref. \cite{AIkedaLCO2016}.
In addition, the upper boundary of the $\beta$ phase is found to be $\approx120$ K.
}

\begin{figure}[t]
 \includegraphics[width = 0.8\linewidth]{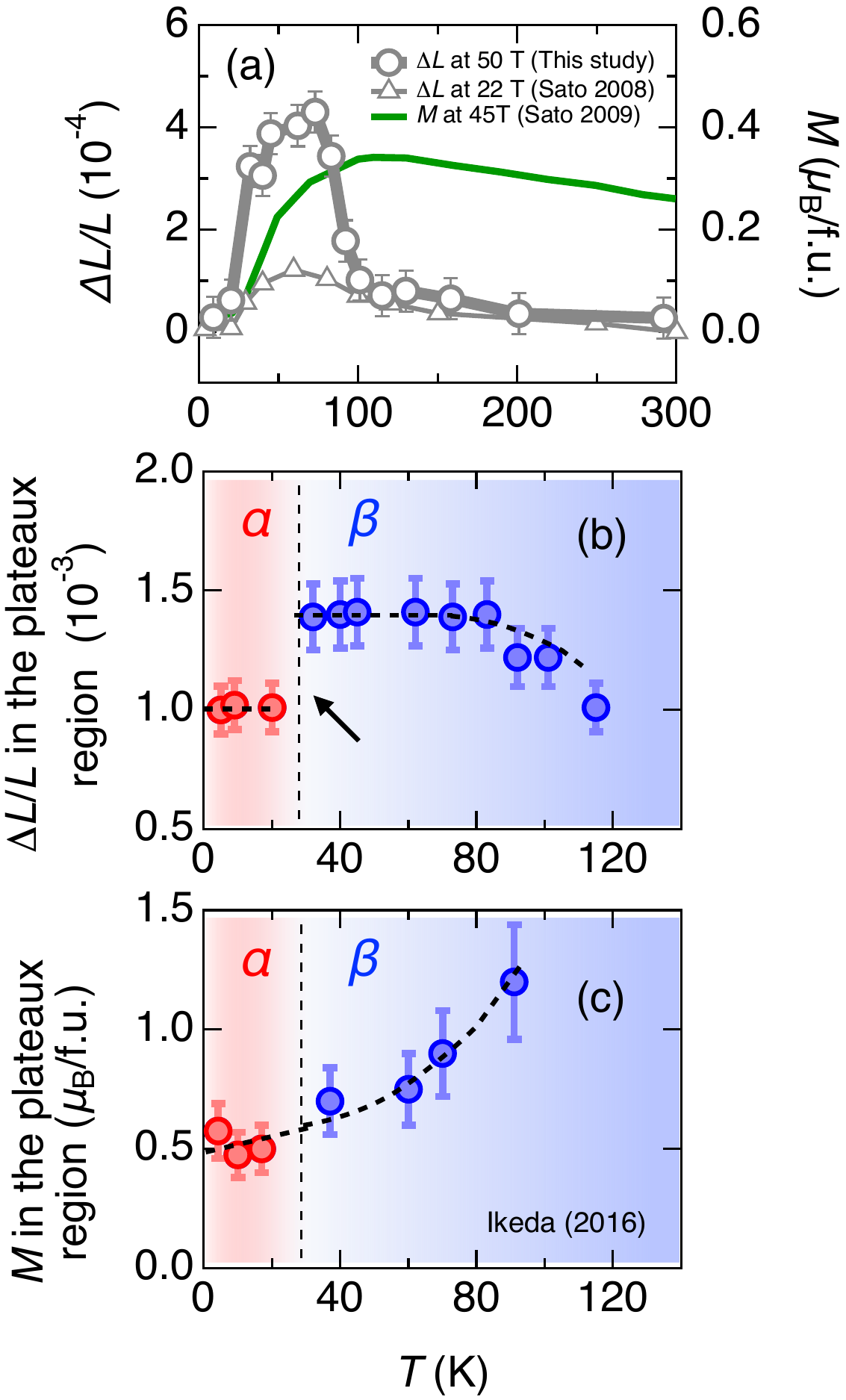}%
 \caption{\textbl{(a) $\Delta L/L$ of $\rm{LaCoO}_{3}$ at 50~T (Present study) as a function of temperature with $\Delta L/L$ at 22~T (imported from Ref. \cite{SatoJPSJ2008}) and magnetization (imported from Ref. \cite{SatoJPSJ2009}) shown for comparison. 
(b) $\Delta L/L$ of $\rm{LaCoO}_{3}$ in $\alpha$ and $\beta$ phases deduced in Fig. \ref{results}(c) as a function of temperature.
 (c) Magnetization of $\rm{LaCoO}_{3}$ in $\alpha$ and $\beta$ phases imported from Ref. \cite{AIkedaLCO2016}.}
 \label{analysis}}
 \end{figure}

\textbl{$\Delta L/L$ is demonstrated to serve as a more direct probe of spin crossover than magnetization [Fig. \ref{analysis}(a)].
$\Delta L/L$ at 50~T shows significant response in the temperature range from 25~K to 100~K and vanishes with increasing temperature at above 100~K, which is consistent with a previous magnetostriction study up to 35 T \cite{SatoJPSJ2008}.
In contrast, magnetization is well susceptible even at room temperature \cite{SatoJPSJ2008}.
These indicate that, at room temperature, the paramagnetic spin moments of the HS or IS are aligned to the external magnetic fields, while the spin-state fractions do not change \cite{note01}.
On the other hand, at low temperatures, both the spin crossover and the spin moment alignment are taking place.
Based on above observations, hereafter in the paper, we regard the change of $\Delta L/L$ as the change of spin state fractions, as a crude approximation.}

\textbl{In Fig. \ref{results}(c), we identify three temperature ranges as $T < 30$~K, 30~K $< T < 130$~K, $T>130$~K, where the $\Delta L/L$ curves are colored in red, blue, and green, respectively.
The former two ranges exhibit phase transitions to plateaux, while the last range shows no transition.
The temperature range $T < 30$~K is characterized with the gapped behavior, while the gapless behavior is observed in 30~K $< T < 130$~K and $T>130$~K.
A striking observation is summarized in Fig. \ref{analysis}(b), where the heights of the plateaux in $T < 30$~K and 30~K $< T < 130$~K are plotted.
It is clear that the plateaux heights are sharply increased by a factor of $\sim1.4$ in 30~K $< T < 130$~K as compared to those in $T < 30$~K.
This asserts that $\alpha$ and $\beta$ phases are distinct phases separated by a first order transition line as is indicated in Fig. \ref{pd} by the horizontal dashed line.
The height of the plateaux in $\beta$ phase decreases at elevated temperatures, which could be a result of the increasing preoccupation of IS or HS with increasing temperature at zero fields.}

\textbl{
Note that single crystalline samples show qualitatively similar results up to 190 T and with sharper transitions as shown in the Supplemental Material \cite{supp1} and in Fig. \ref{ml}(a), whose transition points are also plotted in Fig. \ref{pd}.
However, the polycrystalline data are more advantageous because they are much less fragile against the rapid lattice changes in the transitions.
Note also that the oscillatory features that overlap the plateaux in $T < 30$~K and 30~K $< T < 130$~K as shown in Fig. \ref{results}(a) could result from a shockwave propagating inside the sample as discussed in the Supplemental Material \cite{supp2}.
}


First, we discuss that spin-state crystallizations are favored over excitonic condensations in $\alpha$ and $\beta$ phases based on the present observation that $\alpha$ and $\beta$ phase show $\Delta L/L$ plateaux rather than slopes as shown in Figs. \ref{results}(a) and \ref{results}(c).
With excitonic condensations, spin-states can evolve in a gapless manner with magnetic fields with $a\ket{\rm{IS}}+b\ket{\rm{LS}}$ where $a/b$ can smoothly evolve with magnetic fields \cite{TatsunoJPSJ, SotnikovSR}.
On the other hand, in spin-state crystallizations, the energies are gapfull as a function of spin-state fractions, resulting in spin-state plateaux \cite{TatsunoJPSJ, SotnikovSR}.
This indicates that, macroscopically, the localization dominates over the coherent delocalization of the spin-states in $\alpha$ and $\beta$ phases \cite{note05}.


Second, we discuss the distinct natures of the spin-state crystallizations in $\alpha$ and $\beta$ phases.
For this, we point out the distinct behavior of $\Delta L/L$ and magnetization in $\alpha$ and $\beta$ phases.
In contrast to $\Delta L/L$ showing a jump at the $\alpha$-$\beta$ phase boundary in Fig. \ref{analysis}(b), magnetization is smoothly connected as shown in Fig. \ref{analysis}(c), which is indicative of distinct spin-spin interactions in $\alpha$ and $\beta$ phases.
As representatively shown in Figs. \ref{ml}(a) and \ref{ml}(b), in $\beta$ phase, a linear magnetization \cite{AIkedaLCO2016} and a constant $\Delta L/L$ are observed with increasing magnetic fields.
The behavior infers that, in the spin-state crystal of $\beta$ phase, spin moments are forced to align gradually to the field direction owing to the antiferromagnetic coupling of spins.
This idea of antiferromagnetism is also consistent with the increasing magnetization with the increasing temperature as shown in Fig. \ref{analysis}(c), where, with a paramagnetism or a ferromagnetism, an opposite behavior would be expected.
On the other hand, in $\alpha$ phase, a constant magnetization \cite{SatoJPSJ2009, MoazPRL2012, AIkedaLCO2016} and a constant $\Delta L/L$ are observed with increasing magnetic fields \cite{MoazPRL2012, RotterSR2014} as shown in Figs. \ref{ml}(a) and \ref{ml}(b), and also with increasing temperature as shown in Fig. \ref{analysis}(c).
In the spin-state crystal of $\alpha$ phase, it is inferred that magnetization is saturated with a paramagnetic or a ferromagnetic coupling of spins.

 \begin{figure}[t]
 \includegraphics[width = \linewidth]{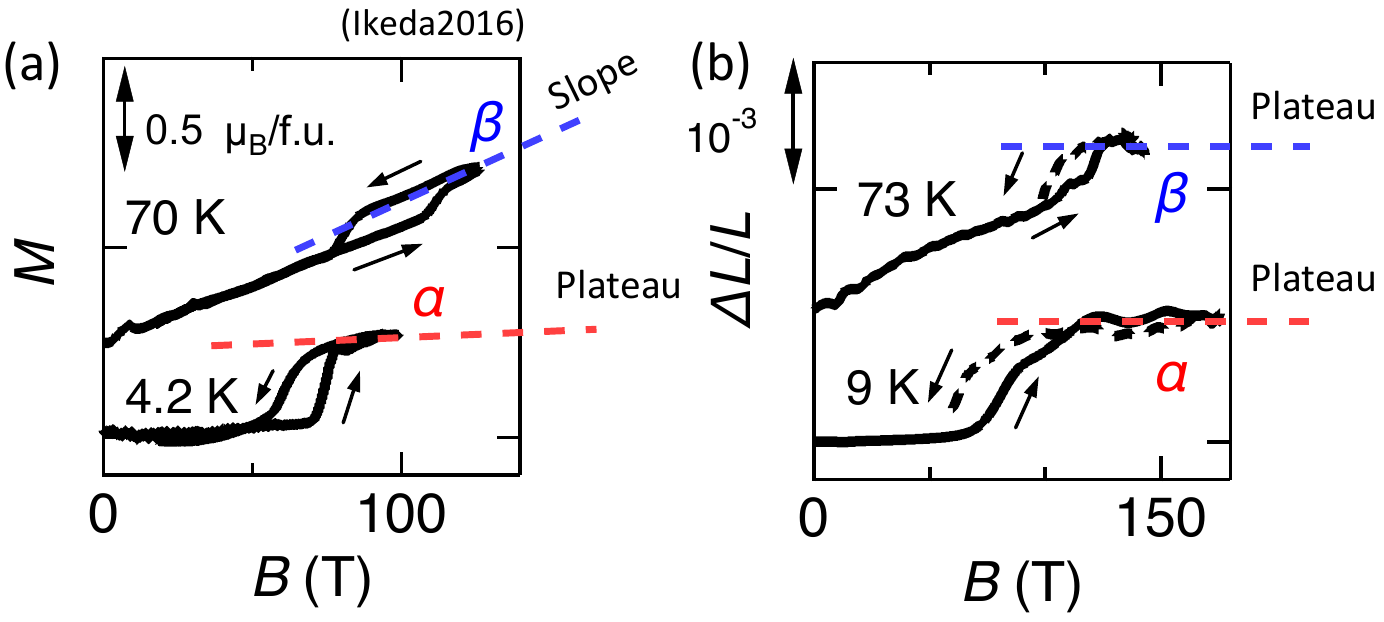}
 \caption{\textbl{(a) Magnetization process of $\rm{LaCoO}_{3}$ up to 130~T at 4.2~K and 70~K. (b) $\Delta L/L$ of the single crystalline samples of $\rm{LaCoO}_{3}$ up to 160~T at 9~K and 73~K, for the comparison of their behavior in $\alpha$ and $\beta$ phases. }
 \label{ml}}
 \end{figure}

 \begin{figure}[t]
 \includegraphics[width = 0.7\linewidth]{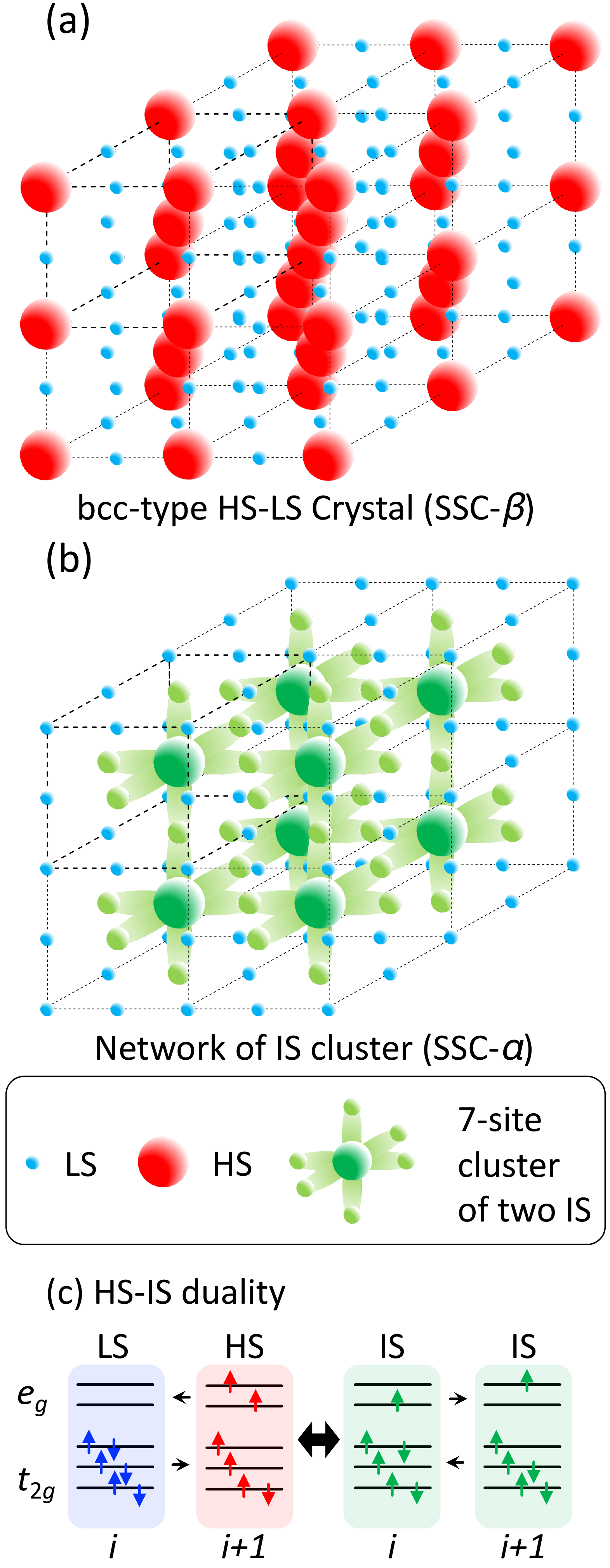}
 \caption{\textbl{
Schematic drawings of (a) HS-LS spin-state crystal forming a bcc-type $2\times 2\times 2$ superlattice (SSC-$\beta$) and
(b) Spin-state crystal of a seven-site IS cluster forming a cubic-type $2\times 2\times 2$ superlattice  (SSC-$\alpha$).
(c) The HS-IS duality. In an seven-site cluster, two IS ions are delocalized.
 \label{ssc}}}
 \end{figure}

Based on above arguments \textbl{and the recent claims on HS-IS duality \cite{TomiyasuArxiv, HarikiArxiv}}, we tentatively propose the structures of the spin-state crystals for $\alpha$ and $\beta$ phases, which are schematically drawn in Figs. \ref{ssc}(a)-(c).
In $\beta$ phase, a HS-LS spin-state crystal with a $2\times2\times2$ superlattice in a body-centered cubic (bcc) lattice (SSC-$\beta$) is anticipated as shown in Fig. \ref{ssc}(a).
In $\alpha$ phase, a spin-state crystal of a seven-site IS cluster with the same superlattice in a cubic lattice (SSC-$\alpha$) is anticipated as shown in Fig. \ref{ssc}(b).
The two spin-state crystals are connected by the HS-IS duality, where the seven-site IS cluster and a HS are dual, whose process is schematically illustrated in Fig. \ref{ssc}(c) in a tight binding view \cite{HarikiArxiv}.
Actually, in a recent calculation \cite{TomiyasuArxiv}, spin-state crystals corresponding to SSC-$\beta$ and SSC-$\alpha$ are shown to become most stable when the lattice is expanded from the LS phase by 2\% and 0.5\%, respectively \cite{note03}.
This is qualitatively in good agreement with the present observation that $\Delta L/L$ is 1.4 times larger  in $\beta$ phase than that in $\alpha$ phase.
Besides, the spin-state crystals in the calculation \cite{TomiyasuArxiv} corresponding to SSC-$\beta$ and SSC-$\alpha$ are shown to have antiferromagnetic and ferromagnetic spin couplings, respectively.
This is also in good agreement with our anticipation for the magnetic interactions in $\alpha$ and $\beta$ phase \cite{note04}.
Further, the observed value of the magnetization in $\alpha$ phase  $\sim0.5$ $\mu_{\rm{B}}$/Co is in good agreement with the saturation of the model SSC-$\alpha$, 0.5~$\mu_{\rm{B}}$/Co with $g =2.0$.
The larger magnetization observed in $\beta$ phase is also consistent with the model SSC-$\beta$.
\textbl{We note that SSC-$\alpha$ under high magnetic fields may resemble the appearance of ferromagnetic phases in epitaxial thin films of LaCoO$_{3}$, in the sense that insulating and ferromagnetic spin-state crystals emerge in expanded lattices \cite{SotnikovSP}.
Various long-range orders are induced by the tensile strains \cite{FujiokaPRL2013, FujiokaPRB2015, YokoyamaPRL2018}, indicating the presence of various high-field phases at the 1000 T range.
Theoretical evaluations considering further neighbor interactions may be needed to verify above arguments.}

\textmg{
Finally, we discuss the temperature change of the sample in the present experiment.
The samples are in the adiabatic condition in the present study because the short pulse duration of $\mu$s does not allow any heat dissipation.
Preliminarily, the sample temperature is measured with an adiabatic condition under millisecond-pulsed magnetic fields up to 65 T \cite{Kihara}.
The results suggest that the temperature changes are non-negligible below 30 K due to the reversible magnetocaloric effect and non-reversible heating at the first order transition, which are of the order of $\pm 5$ K and $+15$ K, respectively.
They are negligible above 30 K with sufficiently large heat capacity \cite{Kyomen2003}.
This indicates that the overall features of our $B$-$T$ diagram are unchanged but that marginal modifications are possible at below 30 K, that could be verified with a hypothetical use of a static 100 T environment [See the inset of Fig. \ref{pd} for a possible modification to the phase diagram].
}

In conclusion, we have reported magnetostriction measurements of $\rm{LaCoO}_{3}$ \textbl{well above 100~T} using a new magnetostriction gauge \cite{AIkedaFBG2017}. 
$\alpha$ and $\beta$ phases beyond 100~T are found to  exhibit $\Delta L/L$ plateaux, which is larger in $\beta$ phase by a factor of 1.4 than that of the low temperature $\alpha$ phase, indicating that $\alpha$ and $\beta$ originate in two crystallizations of spin-states rather than excitonic condensations.
Two models for $\alpha$ and $\beta$ phases are tentatively proposed based on the observed spin crossover and magnetism, considering the duality of HS and IS states.
The field induced spin crossover in LaCoO$_{3}$ is reshaped by electron-correlations leading to the crystallizations of the localized and interacting spin-states in the expanded lattice, thereby suppressing the itineracy of IS and the appearance of excitonic condensations.
\textbl{Considering that the spin-state is far from fully polarized even at 190 T, further emergence of exotic orders at even higher fields of up to $\sim1000$~T are expected with the interplay between the spin state degree of freedom, electron correlations and the further expanded lattices.}

\begin{acknowledgements}
We thank T. Nomura (SPring-8) for careful reading of the text and a valuable discussion.
The work is financially supported by JSPS KAKENHI Grant-in-Aid for Young Scientists Grant No. 18K13493 and the Basic Science Program No. 18-001 of Tokyo Electric Power Company (TEPCO) memorial foundation.
\end{acknowledgements}

\bibliography{lco}

\begin{thebibliography}{51}
\bibitem [{not({\natexlab{a}})}]{note06}%
  \BibitemOpen
  \href@noop {} {} ({\natexlab{a}}),\ \bibinfo {note} {we note that both
  longitudinal and transverse strain should be measured to fully comprehend the
  magnetovolume effect. Such a technical development is underway.}\BibitemShut
  {Stop}%
\bibitem [{not({\natexlab{b}})}]{note01}%
  \BibitemOpen
  \href@noop {} {} ({\natexlab{b}}),\ \bibinfo {note} {note that the
  observations of $\Delta L/L \sim 0$ and $M > 0$ at RT and above 40 T indicate
  that the strain coupled to the spin reorientation is negligibly small in the
  present experiment.}\BibitemShut {Stop}%
\bibitem [{not({\natexlab{d}})}]{note03}%
  \BibitemOpen
  \href@noop {} {} ({\natexlab{d}}),\ \bibinfo {note} {the calculated SSCs in
  \cite{TomiyasuArxiv} corresponding to the currently proposed SSC-$\alpha$ is
  assigned to 1/2-F$_{1}$. While for SSC-$\beta$, 1/2-AF$_{1}$ or 1/2-Ferri are
  assigned.}\BibitemShut {Stop}%
\end{thebibliography}
\end{document}


\title{Supplemental material}
\author{Akihiko~Ikeda}
\author{Yasuhiro~H.~Matsuda}
\affiliation{Institute for Solid State Physics, University of Tokyo, Kashiwa, Chiba, Japan}
\author{Keisuke~Sato}
\affiliation{National Institute of Technology, Ibaraki College, Hitachinaka, Ibaraki, Japan}

\date{\today}
%
%


\maketitle

\section{The results of magnetostriction measurements with single crystalline $\rm{\textbf{LaCoO}}_{3}$}

  \begin{figure}[b]
 \includegraphics[width = 0.7\linewidth]{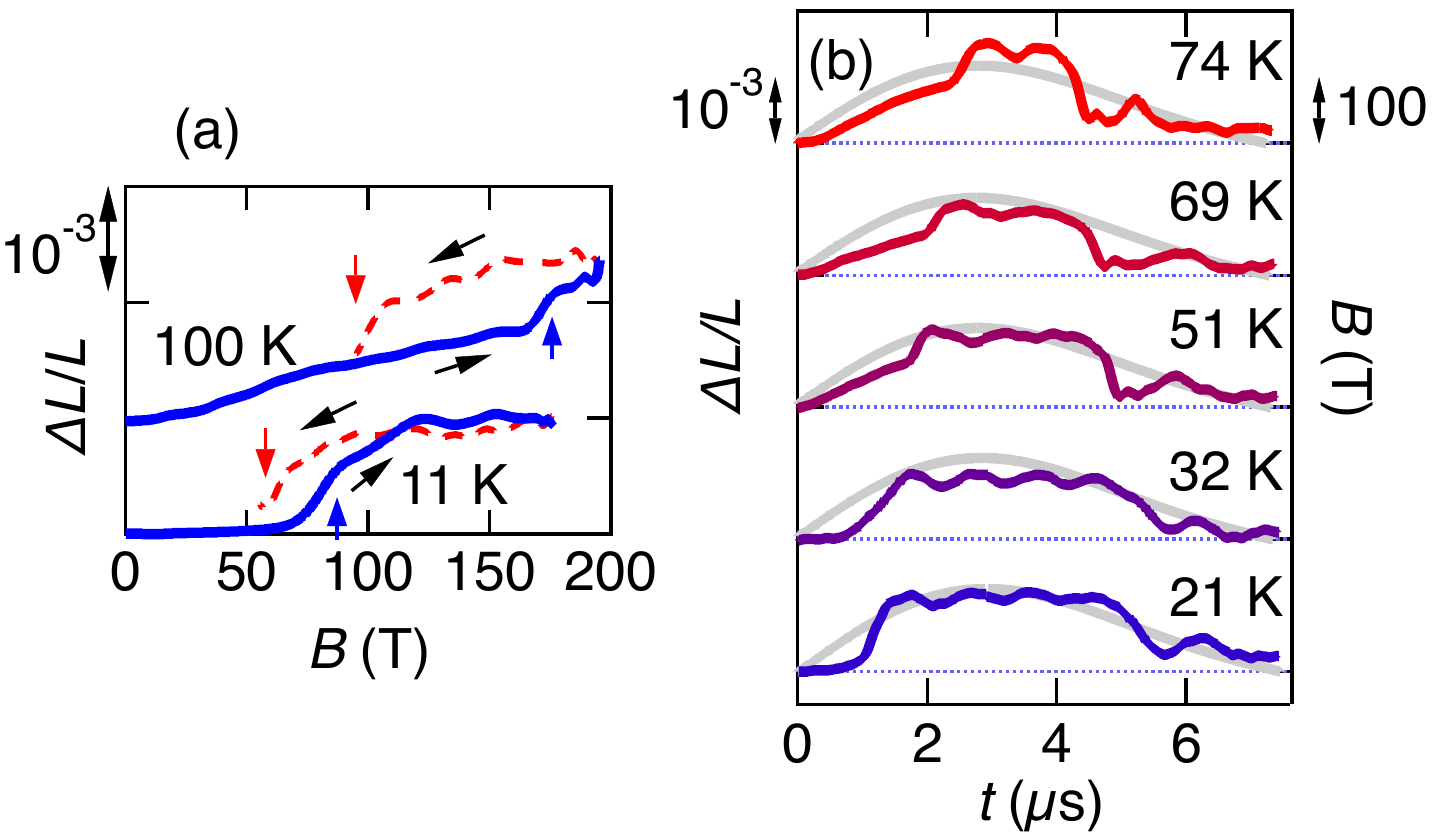}%
 \caption{Result of magnetostriction measurements of single crystalline LaCoO$_{3}$ for various temperatures as a function of (a) magnetic fields and (b) time.
 \label{sc}}
 \end{figure}

Single crystalline samples are measured up to 200 T as shown in Figs. S\ref{sc}(a) and S\ref{sc}(b).
Because single crystalline samples break themselves due to the field induced phase transition taking place too rapidly, it is difficult to observe the temperature dependence coherently with many pulses, being disturbed by a re-gluing a fresh sample to another optical fiber.
On the other hand, polycrystalline samples are robust against the spin-state transitions triggered by the rapid magnetic field pulse.
Poly and single crystalline samples show basically show identical results with a minor differences in the sharpness of the spin-state transitions.

\clearpage

\section{The oscillatory feature in the magnetostriction data}

The oscillatory features overlap the plateaux in (T1) and (T2) as shown in Figs.  S\ref{sc}(b) (this text) and 2(a) (main text), whose origin is considered to be a shockwave propagating inside the sample.
The oscillatory feature after the transition in the down sweep is dumped in (T1) while it is sustained in (T2).
The velocity of the shockwave calculated from the sample size of 2.5 mm and the frequency $\sim0.8$ MHz is $\sim5$ km/s, which is roughly the typical sound velocity of TMO \cite{Naing}.
Deduced $v$ has an error bar of 10 \% which is too large to extract any information on the elastic properties of $\alpha$ and $\beta$ phases.

  \begin{figure}[h]
 \includegraphics[width = 0.5\linewidth]{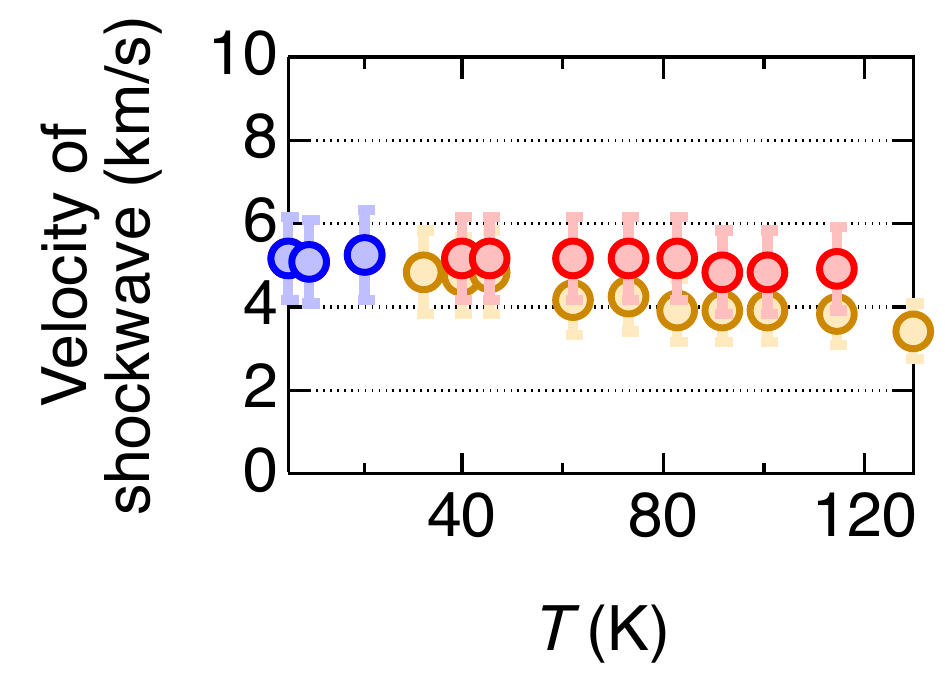}%
 \caption{Analyzed value of the velocity of the shockwave observed in the magnetostriction data in Fig. S\ref{sc}(a) and in Fig. 2(a)
 \label{osci}}
 \end{figure}

\clearpage

\section{Microscopic interactions among spin-states and spins in the proposed SSC$\rm{\textbf{s}}$}

  \begin{figure}[h]
 \includegraphics[width = 0.8\linewidth]{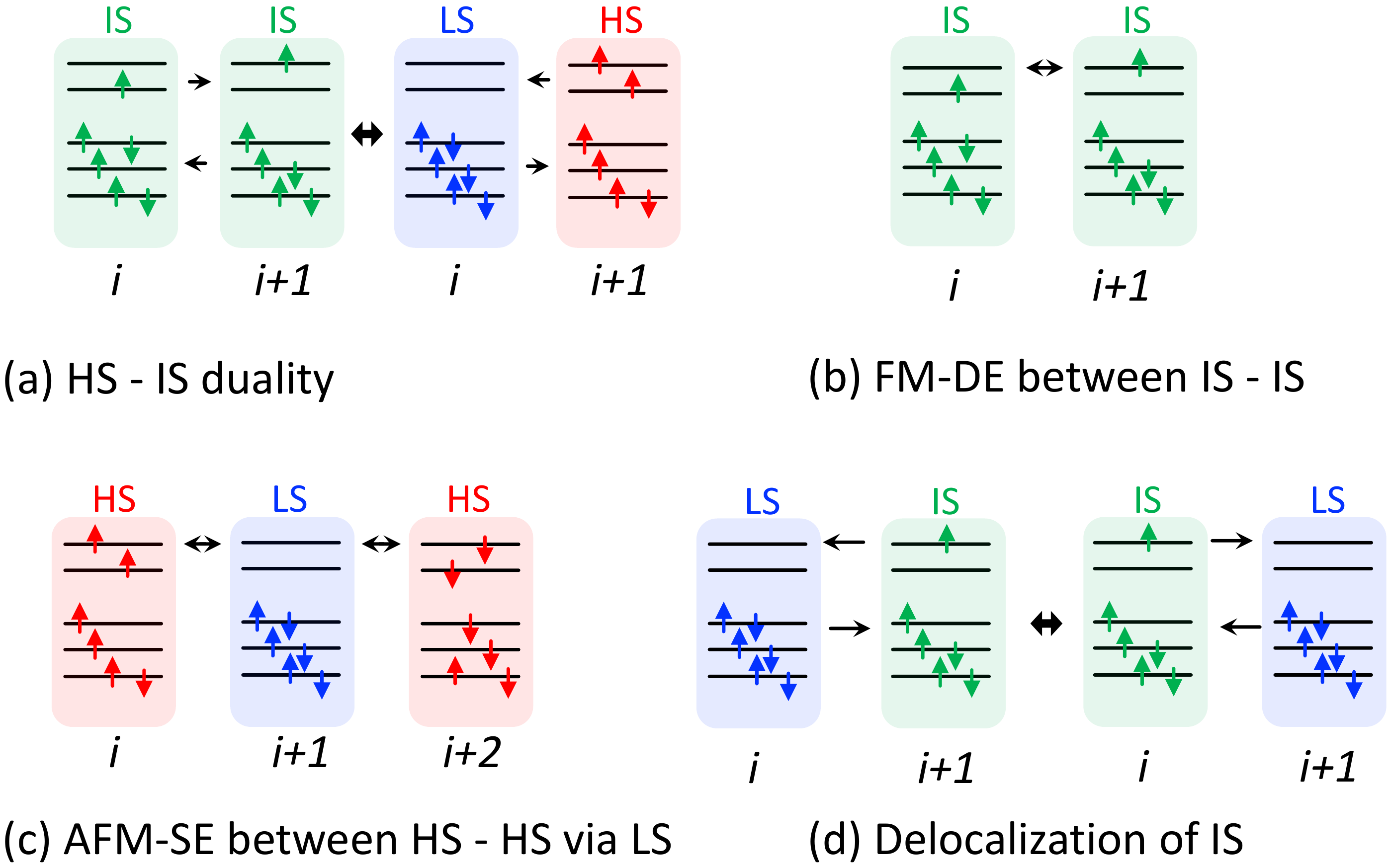}%
 \caption{
 (a) The HS-IS duality. In an seven-site cluster, two IS ions are delocalized as $\ket{\rm{\dots, IS, IS, \dots}}\rightleftharpoons\ket{\rm{\dots, LS, HS, \dots}}$
Schematic drawings of electron hopping process responsible for
(b) Ferromagnetic (FM) interactions of spins between neighboring IS states in a double exchange (DE) scheme,
(c) Antiferromagnetic interactions of spins between two HS sites mediated by a LS site, and
(d) Hopping of a IS state to an adjacent site responsible for the delocalization of IS.
 \label{atomic}
 }
 \end{figure}

Here, we discuss the microscopic mechanism of the inter spin-state interactions and magnetic interactions, which are relevant to the spin-state crystals (SSCs) proposed in the main text.
The HS-LS duality proposed in \cite{TomiyasuArxiv, HarikiArxiv} is schematically depicted in Fig. S\ref{atomic}(a).
The a pair of the electron hoppings in the $e_{g}$ and $t_{2g}$ orbitals transforms a pair of neighboring HS and LS into a pair of neighboring two ISs, and vice versa.
The duality stabilizes the energy of the paired ISs and also the pair of HS-LS because of the itineracy of IS as compared to the cases of the completely localized state \cite{HarikiArxiv}.

In Fig. S\ref{atomic}(b), the microscopic mechanism of the ferromagnetic interaction between neighboring ISs in the SSC-$\alpha$ is depicted.
The double exchange like interactions between the $t_{2g}$ orbitals of neighboring ISs ferromagnetically align the neighboring spins via the Hund's coupling.

In Fig. S\ref{atomic}(c), the microscopic mechanism of the antiferromagnetic interaction between the next-nearest-neighboring HSs.
The 2 step electron hopping from a HS to the next-nearest-neighboring HS via a LS site and 2 more step of returning to the original site act as the superexchange like interactions between the antiferromagnetically aligned next-nearest-neighboring HSs because if the spins are aligned ferromagnetically the 4 step process is prohibited. 
The kinetic energy gain in the propagation of the electron stabilizes the antiferromagnetism.
This may be relevant to the SSC-$\beta$ proposed in the main text.

In Fig. S\ref{atomic}(d), the microscopic origin of the the itineracy of IS is depicted.
The IS adjacent to a LS will be delocalized  between the sites by hopping of the IS between the sites.
The hopping of a IS to the adjacent LS site occurs via a pair of a electron hopping n $e_{g}$ orbital and in $t_{2g}$ orbital.
One IS acts as an itinerant boson on the sea of the LS vacuum in the field theoretical view.
The boson in this case is a spin-full exciton, which may condense or crystalized depending on the hopping parameters in the $e_{g}$ and in $t_{2g}$ orbitals \cite{NasuPRB}.


\bibliography{lco}